# Sustainability concepts for digital research infrastructures developed through ground-level stakeholder empowerment

**Florian Ahrens[1]\*, Dawn Geatches[2], Niall McCarroll[3], Justin Buck[4], Alvaro Lorenzo-Lopez[4], Hossein Keshtkar[5], Nadine Fayyad[5], Hamidreza Hassanloo[5], Danae Manika[5]**

1 International Centre for Island Technology, Heriot-Watt University, Orkney Campus, Stromness KW16 3AN, UK

2 Innovate UK KTN, Unit 218, Business Design Centre, Upper Street, Islington, London N1 0QH, UK

3 University of Reading, Reading RG6 6AU, UK

4 National Oceanography Centre, Joseph Proudman Building, 6 Brownlow Street, Liverpool, L3 5DA, UK

5 Brunel University London, Uxbridge, UB8 3PH, UK

**\* Correspondence:**
Corresponding Author
fja2000@hw.ac.uk

## Abstract

The UK Research and Innovation Digital Research Infrastructure (DRI) needs to operate sustainably in the future, encompassing its use of energy and resources, and embedded computer hardware carbon emissions. Transition concepts towards less unsustainable operations will inform the future design and operations of DRI. A problem remains that, while the skills and knowledge for solving net zero challenges already exist within the UK's DRI community, the mechanisms for sharing them and enabling behavior change are missing. Without adopting community-driven approaches, individual stakeholders may feel isolated and uncertain about how to play their role in the transition. A research programme was funded to give voice to the ground-level stakeholders of the DRI ecosystem for the co-creation of carbon downshift concepts. This article presents the results of the programme, with the goal to inform a fair and just transition from the ground-level, complementing the top-down interventions of energy efficiency policies and renewable energies integration. A workshop-based innovation method was developed for researching stakeholder recommendations and perspectives on the sustainable transition of the UK's DRI. We find that giving a purposeful voice to the stakeholders for shaping their own future sustainable DRI environment can be achieved by a guided, expert-integrated, interactive and problem-focused workshop series. The chosen workshop design is impactful on creating bottom-up agency for climate action by first defining the high-level problems of unsustainability in energy and fossil-fuel consumption, and then connecting them to the ground-level circumstances of DRI stakeholders. This approach to stakeholder management should initiate a sustainable transition that promises to kick-start impactful changes from within communities, adding to high-level efforts from economics, policy, and governance.

# 1 Introduction

Information and communication technology (ICT) account for 4 - 6% of the global electricity consumption and estimations of the share in global greenhouse gas emissions range between 2% and 3% (Ross and Lorna, 2022). The international research community makes extensive use of ICT in the form of high-performance computing (HPC), data storage, databases, or computers and digital communication in general. The UK Research and Innovation council (UKRI) is responsible for extensive use of ICT; UKRI is the national research and innovation system comprising seven research funding councils: Science and Technology Facilities Council – STFC; Biotechnology and Biological Sciences Research Council – BBSRC; Engineering and Physical Sciences Research Council – EPSRC; Economic and Social Research Council – ESRC; Medical Research Council – MRC; Natural Environment Research Council – NERC; and the Arts and Humanities Research Council – AHRC, Research England, and Innovate UK. In total UKRI employs about 8,000 people and the majority of whom use a form of digital asset to deliver their work across the arts, humanities and science sectors. The collection of the digital assets is described as the Digital Research Infrastructure (DRI), and the employees (and therefore users in *any capacity,* e.g. developers, users, administrators, access enablers etc.) of these UKRI DRI assets are described as its stakeholders. As well as data centres which provide data storage and computational capability, the DRI also provides an encompassing ecosystem of software and code libraries and most importantly, a user community which enables the sharing of knowledge and expertise amongst all stakeholders. UKRI has set the target to achieve net zero carbon emissions from its DRI by 2040 (Juckes et al., 2023). To stay in line with climate science, net zero goals must translate into a downshift of fossil-fuel consumption in human activity systems by 80% in the coming decade (Calverley and Anderson, 2022). This means there is a need to reduce emissions drastically from all scopes, ranging from the operational to the embodied carbon. Apart from greenhouse gas emissions, the ICT sector potentially risks transgressing other planetary boundaries through the polluting upstream practices of the mining and the extraction of raw materials (Pohl et al., 2021). Sustainability is the capacity to continue performing essential activities into the future without risking the satisfaction of essential need and the degradation of finite and renewable resources. To put this into the context of UKRI DRI, a sustainable transition to net zero emissions means to drastically reduce carbon emissions in its operations, but also to purposefully re-design its institutions and processes for safe and sustainable operation in the future. Sustainable UKRI DRI includes a community culture of frugality when operating the various DRI facets, the downshift of fossil-fuel/non-renewable derived electricity for powering and cooling components, and a reduction of hardware-embedded fossil fuels.
Proposed interventions to reduce greenhouse gas emissions include deploying more renewable electricity, and increasing computational and energy efficiency (Awasthi et al., 2019). The material footprint of the ICT hardware could be reduced through sustainable procurement, increased lifetimes, and recycling towards less linear economies of waste (Huang et al., 2020). For UKRI DRI,  current net zero plans include renewable electricity and energy storage, increased hardware and software efficiency, improved utilization through multi-use applications, reduction in hardware embodied carbon, carbon accounting, and carbon capture methods (Juckes et al., 2023). However, the real-world impact of these measures is yet unclear. Renewable energy scenarios pose economical (King and van den Bergh, 2018), environmental (Slameršak et al., 2022), and raw material demand issues (Kalt et al., 2022; Sovacool et al., 2020; Watari et al., 2021). The application of energy storage into renewable energy systems intensifies the aforementioned issues. While energy storage will be essential, the demand for it should be reduced as much as possible through re-designing energy use systems towards higher, direct self-consumption (Desing and Widmer, 2022). Increasing efficiencies on multiple levels can eventually saturate (Murphy, 2022) and risk producing rebound effects without establishing climate-safe energy and resource constraints on a systems level (Brockway et al., 2021). Historically, rebound effects were observed when increased efficiency frees up resources that are then invested into growing the enterprise instead of applying strategic foresight, thus not reducing but increasing environmental impacts.

Including the  stakeholders of UKRI DRI  in a bottom-up transition process will enable the DRI community to participate in the ground-level re-design of their own climate safe and sustainable socio-technological-economic future digital research environment (Shaw et al., 2009). Many of the skills and knowledge that will be needed to solve net zero challenges already exist within the UKRI DRI community (Juckes et al., 2023), but mechanisms for sharing them and enabling behavior change are missing. Without adopting community-driven approaches, individual stakeholders may feel isolated and uncertain about how to play their role in solving these complex problems. Stakeholders across the UKRI DRI community are currently lacking ways for their voices to be heard and need support to take their first practical steps on their net zero journeys. Without the stakeholders' inputs, UKRI DRI will struggle to discover where future investment should be directed to develop new technologies, business models, and policies for reducing unsustainability in their DRI processes.

The scope of this work is to report on the development a workshop series for finding ways to connect and empower the UKRI DRI community to meet the net zero challenge. Co-designed projects contribute to consensus creation, build trust between different stakeholder groups, and can transform behavioral, policy and social outcomes (Balvanera et al., 2017; Reynolds-Cuéllar and Delgado Ramos, 2020). The workshops were designed for UKRI DRI stakeholders to take ownership of their path to net zero. The stakeholders invited were from all nine UKRI councils: computing infrastructure providers and experts; application developers; (scientific and business) users of the software and/or HPC resources; funding and data providers; users and providers of digital repositories; and technology designers.

The main steps of a bottom-up sustainable development process are community building, problem definition, co-production of solutions, and definition of projects (Lang et al., 2012). To initiate such a process for UKRI DRI, the research focus of this paper is "How to design a participatory workshop series for UKRI DRI stakeholders to define community-led, key net zero fossil-fuel downshift actions". Our work feeds into the wider efforts undertaken by UKRI Net Zero DRI on policy design, efficiency improvements, carbon counting, decision making, case studies, asset mapping, and behavior studies (Juckes et al., 2023).

This article demonstrates that the chosen workshop design creates stakeholder agency for DRI climate action by first defining the high-level problems of unsustainable energy and fossil-fuel consumption, and then connecting them to the ground-level DRI work environments. The participants self-reported an increase in actionable knowledge for achieving net zero in their DRI environment over the course of the workshops. Technical recommendations are synthesized based on the co-designed sustainability concepts. We conclude that this way of stakeholder management in sustainable transition processes could promise to kick-start impactful changes from within communities, complementing top-down efforts from economics, policy, and governance.

The article commences describing the chosen workshop design. The technical recommendations from the workshops and the self-reported participant survey data are laid out in the results section and are then discussed. Concluding remarks summarize the article and give a forward-looking perspective to future research based on the learnings.

## 2  Methods

Devising engaging online workshops is a deceptively complex and time-intensive activity; although the online aspect of the workshop is straightforward in so much as participants need access to the internet and a communication platform. The lessons learned from devising and running the workshop series are summarized below, and a fully descriptive workshop toolkit is available in the supplementary material. The toolkit summarizes the learnings of how to run online workshops and includes further details.

Three sequential workshops were held over a period of four weeks in October 2022. Participants were invited through an online expression of interest (EOI) form with relevant information about the workshops using a snowball dissemination technique within the UKRI DRI community, and the professional and personal contact lists from across UKRI research councils. Snowball dissemination was deemed to be the most appropriate method for participant recruitment as it enabled reaching groups of people relevant to the UKRI DRI context outside of the researchers' professional networks. Although snowball dissemination was advantageous for reaching a multidisciplinary audience, there are drawbacks such as the difficulty to generalise beyond the sample studied. Therefore, this study does not aim to provide a generalised statement, rather the contextual knowledge of a multidisciplinary audience. The invite and EOI was also distributed using mailing lists including the UKRI mailing lists and the American Geophysical Union Informatics Google group (~800 members). Organisational social media was used, including those of UKRI, NERC, NOC and BODC, with followers across the accounts totalling at least 2,000. The EOI was used as a means of capping workshop numbers in case the number of applications exceeded the number of workshop places. Had this been the case the participants would have been selected using their EOIs. However, fewer applications than spaces available were received and hence no selection methods were used. Recruitment was hampered by the government communication in the wake of the passing of Her Majesty Elizabeth II, but each communication channel was used to advertise the workshops on at least one occasion.

Incentives were offered through three donation or voucher choices: UNICEF[1] donation (altruistic and humanitarian), Ecologi[2] donation (altruistic and climate-focused), or lifestyle[3] voucher (egoistic and general). The 4-hour long workshops were held online on the Zoom platform. Each workshop was recorded and transcribed. Breakout rooms were used to increase the participant engagement with smaller group sizes. Expert speakers were invited to stimulate knowledge co-production. A two-week gap between the workshops with before- and after-workshop online surveys ensured that the participants could reflect on their experiences. The surveys probed the participants' perceptions about comfort levels in engaging with the UKRI DRI community, the importance of net zero for UKRI and their job role, their knowledge on the topic of net zero DRI, and their control and intentions to take action to meet net zero targets. The collected data from the surveys and the transcripts were subject to an approved research ethics proposal. Ethical approval of the research was granted by Brunel University London. The participant data (survey results, workshop transcripts, submitted action plans) were anonymized.

[1] Relief agency for children

[2] Crowdfunding corporation

[3] Amazon gift card

## 2.1  Participant information

The participant information was gathered in a pre-workshop 1 survey that came with the EOI. Most of the participants were associated with NERC, STFC, and EPSRC, comprising stakeholders mostly from three of the nine organisations of UKRI. Their interests were in data centers, software development, HPC infrastructure, training skills, and hardware. The participant group consisted of 14 males and 10 females, mainly 41 to 50 years old, with no participants younger than 21 or older than 60. Most participants held a PhD, and are active in a senior role in their organisation. The geographic location of most participants was the UK with some being located in France, China, or Poland. Detailed information about the participants of workshop 1 can be found in Table 1 in the appendix.

## 2.2  Workshop Series Overview

### 2.2.1  Workshop 1

Workshop 1 addressed the characteristics of DRI energy consumption and ways to increase energy efficiency. The different breakout rooms were guided to discuss the participants' own impact on the energy consumption of their organization and how the perceived control is over energy efficiency measures. An emerging topic was the discussion of how energy consumption could be monitored for generating the necessary data to base efficiency measures on. The workshop developed an understanding of the term 'net zero', how it relates to the DRI, and the role of energy consumption in the concept of 'net zero'.

### 2.2.2  Workshop 2

Workshop 2 addressed incumbent fossil fuels in the DRI system and their associated carbon emissions, considering scope 1, 2, and 3 emissions. It was argued that, after implementing energy efficiency measures, there would still be a certain amount of embedded fossil fuel left in the system. This "leftover" fossil fuel needs to be downshifted to meet the high-level targets of climate science. The workshop examined behavior change mechanisms within current designs of the DRI system to achieve the climate-safe limit on system-wide fossil-fuel consumption. The discussions in the breakout rooms were facilitated toward how to identify the carbon intensities of the respective facilities and what behavior change measures would impact fossil-fuel consumption. A cap-and-share system was discussed to introduce the concept of bottom-up constraints on fossil-fuel consumption and the maximization of renewable energy in a future, climate-safe DRI.

### 2.2.3 Workshop 3

Workshop 3 related the discussions of the prior workshops back to the ground-level circumstances of the participating DRI stakeholders by addressing key actions to meet net zero targets over the next 5 to 15 years. The participants were asked to develop net zero action plans with aims and objectives, with respect to the feasibility of their proposed actions and the transferability of actions across research communities. The participants were empowered to develop their personal action plans by responding to five 'W' questions (see figure 1). Key policy changes were also addressed in the action plans. The rationale for this workshop was to give voice to the UKRI stakeholders by developing concrete, ground-level recommendations for the UKRI council.

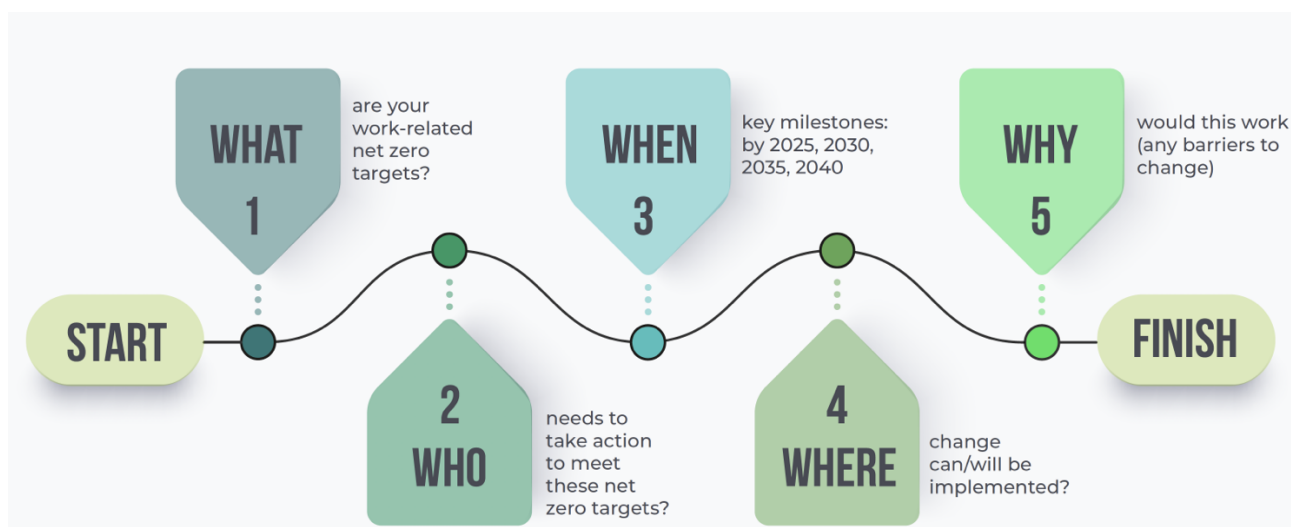

Figure 1 The action plan framework based on 5 W-questions

## 3   Results

The results section lays out the qualitative results from the commissioned net zero action plans, the workshop transcripts, and the quantitative results based on the participant before- and after-workshop surveys.

### 3.1   Planned actions to meet net zero from various key stakeholder perspectives

Action plans for achieving net zero were developed within the workshop series (see supplementary material). Points were made on implementing a carbon-aware strategy for hardware replacement, implementing IT infrastructure for using power monitoring, presenting power consumption data to users for spreading awareness of their HPC jobs, using Research Software Engineers (RSE) to review codes and workflows, and to provide HPC users with carbon estimates to help them understand their emissions and highlight the importance of reducing their emissions. The action plans are synthesized in the 5-W question framework below.

***What** are your work related net zero targets?* This question revealed the range of net zero actions based on the occupations of the attendees, and included supporting end-users of DRI to monitor their own carbon footprint by providing carbon-monitoring tools, optimizing software, and introducing carbon-aware procurement practices. The concept of wider community engagement, building collaborations and creating Net Zero Champions were common themes shared across action plans.

*Who needs to take action to meet these net zero targets?* By answering this question, the participants were broadening the vision of all the possibilities of utilizing human resources and management to achieve mutual goals towards net zero. There was consensus across the action plans in that there is no one person solely responsible for meeting net zero targets. Action needs to be taken at all levels by, for example, end-users of UKRI DRI, researchers, infrastructure managers, executive level directors of the research councils, deans, and vice presidents of universities. The importance of effective communication strategies also featured as a common theme.

*When will these net zero targets be achieved?* The intention of this question was to encourage the participant to draft a set of key milestones based on the short, medium, and long term for their specific net zero action targets. The inclusion of time meant considering the order in which net zero based activity is undertaken, and how capacity can be built for further action. Most action plans contained immediate actions i.e., implemented from now (2023) and completed by 2025, and these focused on the gains made from harvesting low hanging fruit such as community-building, optimizing current software and building new carbon accounting tools. The second common milestone was 2030, in the short to medium term from today's date (2023) and included actions such as aligning targets with an organization's business plan for redevelopment, embedding green procurement practices and building DRI stakeholder engagement in behavior change.

*Where can or will these changes be implemented?* This question revealed the extent to which participants recorded the reach of their activities across their organization, the different sectors, or broader companies and partners, making visible the scope of these changes and how they could affect net zero goals. The common response was local implementation within teams, groups, and DRI communities. Infrastructure was also mentioned insofar as providing a service for DRI users, such as tools for carbon accounting and data storage, i.e., tools to empower users with net zero choices. There was also a common reference to the creation of guidance and policy both within UKRI, across sectors, outside UKRI and by the government.

*Why would this work and are there any barriers to change?* By asking the participants why they think this could work, they were encouraged to think about the opportunities that could help achieve net zero, all the possibilities, and the limitations and barriers they envisage affecting their path to net zero. A theme shared across the action plans was the ability to harvest low-hanging fruit that involve a concrete set of actions, and demonstrate the impact of incremental net zero gains. Clear demonstrations would help educate their peers and wider communities, which in turn builds consensus and momentum, in readiness to tackle more challenging and ambitious net zero goals. The barriers included time and funding pressures to enable communities to explore and adopt new behaviors, learn new skills, and build new tools. Without top-down leadership backed by funding, some participants anticipated difficulties in reducing the resistance to change that could prevail from the ground upwards. A further barrier originates externally such as the availability of carbon data from suppliers, e.g., the suppliers' own transparency of the environmental footprints of their products. Overall, a combination of community-building with gradually implemented policy enforcement would be successful to achieve net zero goals, such as all research proposals requiring an environmental sustainability assessment.

## 3.2  Recommendations based on stakeholders' engagement

Many of the workshop discussions came up with suggestions or recommendations to help with the net-zero transition that fell into one or more of the following themes:

**Operational** - these recommendations cover the way that data could be stored in a more carbon-efficient way and shared (re-used).

**Resource Monitoring** - the workshop participants identified the need to develop tools and systems to monitor electricity and carbon usage across DRI.

**Community and Behavior** - during the workshops a common theme emerged around behavior change and encouragement, which could be more successful with the support of a community.

**Procurement and Funding** - these recommendations arose from the workshop discussions around the organization of project funding to encourage net zero related efficiency improvements.  On a related topic, the DRI equipment maintenance and procurement processes could be adjusted to help reduce the DRI carbon footprint over time.

As expected, some of the topics discussed related to highly technical issues, for example in the areas of resource monitoring or efficient methods for scheduling jobs in HPC systems.  In some cases, the discussions covered opinions that were obstacles to the transition, for example:

> "So what's the point in doing anything to [a] HPC facility? Instead, you should just ground 2 transatlantic flights a day and you've saved more than the whole cost of UKRI DRI combined by a massive amount."

While the absolute amount of emissions from different sectors is important to consider in the transition, this statement is an example of a false dichotomy, in that it will be essential to reduce the carbon footprint and intensity of *both* the airline industry *and* the UKRI DRI. Even with potentially smaller emissions from UKRI DRI compared to the airline sector, not acting will create risks of continuing the unsustainable business-as-usual activity in the future.

The participants generally agreed that behavior change was going to be difficult and required community support, but individuals could see ways they could take action to help:

> "Would you spend a lot of time rewriting your code? Um sorry to sound so cynical. I'm not so. I think you have to change the incentives and motivation."

> "So, I think it's difficult, because I think there's a proportion of [the] community that gets it, and will be on board, but that is, those are not the people that we need to convince, right?"

> "We might not be able to control how the data centres power themselves. But we will be looking at how we can encourage our scientists to run code more efficiently, to be able to monitor their code more efficiently."

> "Especially some students sometimes don't optimise their code at all. They just kind of release the wild beast into space and it just takes all the resources, and the System Administrator sends e-mail. Why did you do it? Never do this again, you know?"

Clearly, there are already some good practices being established, suggesting that solutions already exist within the community but are not yet evenly distributed.

> "We got a team of researchers, and they monitor every single job for its efficiency, and they graph it and look at it on a weekly basis. Anything that's under a certain percent of efficiency,  they'll go off and talk to the researchers and help them optimise the code to bring that back up again."

One recurring operational/community suggestion was to promote the sharing and re-use of data, so that computationally intensive workloads could be avoided altogether.

"My main job is to look after an online portal, and that hosts metadata from lots of different organisations. The idea being so it's all about sharing the data and having the same standards, so that we can all reuse data as much as possible, which obviously has its carbon emission benefits itself."

Expert talks highlighted the emissions embodied in the manufacture of HPC equipment, prompting some discussion on the need for careful procurement and management of the lifetime of equipment.

"For me the embodied carbon emissions is quite striking actually, and that does highlight the need to think carefully about making good use of solutions that we have in operation"

## 3.3  Survey results & analysis

The pre- and post-workshop surveys showed that the workshops did not influence intentions to take action. Based on the 8 participants that fully completed the series and surveys, we found that the sequential workshops were effective in advancing knowledge, but not intentions to take action. More specifically, we found that knowledge improved from the pre-workshops survey to after the sequential workshops. This indicates that sequential workshops as a community engagement mechanism can help DRI stakeholders build and advance knowledge, which in turn could influence their intentions to take action to meet net zero targets (based on prior research knowledge having the capacity to  influence behavior - but also there is caution against assuming that knowledge always leads to behavior change (De Meyer et al., 2021)). The survey results are presented in table 2 in the appendix.

# 4  Discussion

## 4.1  Technical discussion

During the workshop the participants collectively made several informal, qualitative recommendations, many of which were technical in nature, forming part of the results described in section 3.2.  Formalizing and formally agreeing these recommendations was not a goal of the workshops but the recommendations were documented in the project report (Manika, Danae et al., 2022) and passed back to the UKRI Stakeholders. These ultimately contributed to the UKRI Net Zero DRI scoping project final technical report (Juckes et al., 2023) which synthesizes a number of toolkit themes (sets of related approaches and tools useful for transition engineers) from the participant recommendations and from results originating from the other sandpit projects. The technical discussion will relate the developed recommendations (see section 3.2) to the final technical report issued by the UKRI DRI scoping project.

Some detailed **technical and operational recommendations** focused on the data used by the research councils. The development of sophisticated online portals for efficient data storage, sharing and re-use was proposed, in alignment with the broader objectives of  open source and FAIR (findable, accessible, interoperable, and reusable) data (Wilkinson et al., 2016).  Re-use of FAIR data where possible was considered by participants to be particularly important to avoid the energy costs of unnecessary re-computation. Another operational recommendation considered the use of green schedulers to favor running computation when the grid was able to provide energy from renewable sources. Green scheduling of different types of computing workload was also investigated by the HPC-JEEP and the Value sandpit project (Boulton, 2023; Turner and Basden, 2022). Both recommendations were incorporated as "Action-based research" and "Green software engineering" toolkit themes.

**Resource monitoring recommendations** considered the need to develop better tools and systems to monitor electricity and carbon usage across the DRI. This was thought to be essential for properly quantifying improvements in software and hardware efficiency. Tools for monitoring energy and carbon consumption were explored by the IRISCAST sandpit project (Hays et al., 2023). These contributed to the "Green software engineering" toolkit theme.

**Community and behavior recommendations** also emerged from the workshops. A common theme emerged around the difficulties of behavior change. Effective research software engineer training and community building can help researchers to overcome these challenges. A green computing database will help communities to share knowledge. These recommendations appear in the final report mainly under "Knowledge Hub" and "Recognise Shared Responsibility" toolkit themes.

Several recommendations were addressed towards new policies to assist with the net zero transition. Some participants envisaged the inclusion of carbon/energy budgets in project proposal assessments. Funding to encourage training, community building and community-led knowledge transfer activities was also recommended.

**Procurement and funding recommendations** included the fine tuning the procurement, maintenance, refurbishment and disposal of equipment (Bashroush et al., 2022) to take into account the embedded carbon cost of their manufacture as well as their energy consumption. Cap-and-share policies for managing computing workloads were discussed in the workshops and considered to be important. This group of recommendations contribute to the toolkit themes "Recognize shared responsibility", "Mission Focus" and "Working with peers and suppliers".

The recommendations from the workshops were necessarily broad and imprecise in a technical sense. A series of follow up workshops with a more technical focus could be designed to explore some of the technical recommendations discussed above more deeply. These workshops could bring together participants with particular expertise and interests to focus on specific technical challenges with the goal of reaching agreement on more detailed recommendations to pass back to the wider community.

## 4.2 A vision for sustainable net zero DRI

The recommended actions and discussions of the workshop participants are interpreted as "a day in the life of" a DRI user in the near future. This future vision is captured in box 1 below. Business-as-usual forecasting methods usually extrapolate current trends in the future, risking the continuation of unsustainable practices (Irwin, 2018). Visioning on the other side gives the unique opportunity for path-break to imagine a sustainable future where identified problems have been resolved.

Box 1 "A day in the life of" a UKRI DRI user in the near future - a story synthesised from the imagination of workshop participants.

Eliza arrives at the department at 8-30am. Her day starts with a meeting with several other scientists and with two of the research council's green research software engineers to catch up with progress on a project to develop a new dataset. They check the current energy usage on some initial processing, collected by the research cluster's advanced energy monitoring systems. The figures show they are well within the energy-carbon budget agreed with their research council for the project. The meeting concludes with a review of the source code for the next stage in the project. A few areas for potential improvement are identified, but no glaring inefficiencies were found - all scientists and engineers are incentivized by the research council to spend time training to code efficiently and belong to an active community that exchanges ideas to maintain and improve code efficiency and re-use.

After the meeting, Eliza's morning coffee is interrupted by an alert from the research cluster's green scheduler. A few of the experimental simulations that she had submitted the previous day have been delayed due to the shortfall in renewable energy entering the electricity grid. The scheduling system's forecasting feature tells her that the simulations are highly likely to be scheduled within the next 24 hours and Eliza is reassured. In the meantime, she can start looking at the results from the completed simulations. The green scheduling policies adopted by the research cluster matured over 10 years ago, incorporating suggestions from the user community, and everyone is happy with the occasional delay, because they know that the system is fair. In this atmosphere of cooperation, research cluster administrators can spend most of their time monitoring and maintaining the system, ensuring that hardware lifetimes are maximized - everyone is acutely aware of the embedded carbon footprint of new systems.

After spending the rest of the morning in a teleconference with collaborators on a paper, and then a pleasant lunch break with local colleagues, Eliza begins work on a new research proposal. She can quickly survey the existing datasets available via her research community's advanced research data catalogue. Developed and refined over 20 years, this system provides the reliable storage of, and access to, datasets developed and maintained by hundreds of researchers. Eliza is able to quickly identify some relevant data that is suitable for reuse, and notes this in the research proposal, knowing that this will lower the project's processing demands and be appreciated by the proposal's reviewers. A second catalogue allows Eliza to quickly find software tools (that are already certified to be energy efficient) that will form nearly all the building blocks she'll need for running the project's experiments. Eliza then has a brief exchange of messages with the research cluster's administrators to prepare an energy-carbon budget for the proposal. With this out of the way, Eliza can spend the rest of the afternoon developing the more interesting, scientific aspects of the proposal.

## 4.3  Reflecting the method

The most important question to answer first is 'Why run a workshop?' The time, energy and commitment required to run a workshop needs to be justifiable to the event organizers and the proposed attendees. Thinking carefully about this question might lead the organizers to propose a different activity that meets the intended outcomes. Assuming a workshop is the best format, thinking about its structure, e.g., a single or multiple event(s) leads to considering the time commitment required from participants and organizers. This leads to thinking about the availability of the participants, e.g., might they share similar availability, or would the workshop need to run during conventional, non-working hours? Then follows consideration of the workshop content, which needs to be built around the intended outcomes, i.e., what needs to be undertaken to achieve the desired outcomes? The phenomenon of online fatigue induced during the Covid pandemic by numerous online meetings that became the norm across many organizations - including UKRI, can be mitigated by planning dynamism into an online workshop. For example, the content and structure can drive engagement by including a variety of domain specialist speakers, including interactive sessions and breaks of sufficient length to allow participants to step away from the screen and return refreshed. It is important to run an online workshop using resources that are widely available and familiar to the targeted participants, because this reduces the time wasted on technical issues during the workshop. The number of attendees needs careful consideration too because this determines the number of organizers and their roles during the workshop itself, depending on the proposed activities, e.g., breakout rooms need convening, discussions need facilitating, and notes may need to be taken. Practicing the running of a workshop helps ensure it runs smoothly on the day, e.g., test the links in the joining instructions, test the timings, accessibility of resources etc. Finally, carefully targeting the recruitment of participants increases the chances of running a successful workshop because the experience of participating will reward the attendees in addition to achieving the intended outcomes.

### 4.3.1 Rationale for the workshop development

Design is used to "[change] existing situations into preferred ones" (Simon, 1996), making this workshop series part of a whole-system process to redesign the existing DRI environment into a state of sustainability. Design process starts with the definition of the problem and the associated stakeholder needs, followed by the conceptual and detailed design phase, and finally the implementation phase (Howard et al., 2008). The workshops can be understood as the first step of the sustainable re-design of UKRI DRI as they focus on the present problems, the needs for, and in, UKRI DRI, and potential future concepts. Real-world sustainable transition projects require collaboration of various stakeholders (Schneidewind et al., 2018). The workshops aimed to achieve stakeholder engagement through a range of diverse stakeholders from different DRI backgrounds. The creative co-production of scenarios was facilitated by the problem-focused design of the workshops (Dorst and Cross, 2001).
The workshops received generally good feedback for their planning and execution, and the participants found the workshops practical, inclusive, and thought-provoking, validating the research team's planning efforts. Based on the participant feedback and the authors' reflections, the workshop design achieved a rigorous problem understanding (the role of energy and carbon in UKRI DRI), presented expert contributions on recent developments, and gave voice to the UKRI DRI stakeholders through individually developed action plans. Therefore, 3 main achievements of the workshops are the increased knowledge of the participants as evaluated in the surveys; the developed action plans; and the recommendations. But do increased knowledge and intention lead to action? Cognitive sciences hold against this by arguing that increasing awareness of climate issues does not necessarily lead to corrective agency of incumbent stakeholders (De Meyer et al., 2021). This can be seen in the participants' self-reported medium levels of control over achieving net zero in their own job. Nevertheless, change comes about if knowledge and intention are actionable from within the ground-level circumstances of stakeholders, and at the higher level of decision makers. The remaining question therefore is, what do the next steps need to be for realizing the plans and intentions into action?

# 5  Conclusion

The UKRI DRI Net Zero Scoping Project enabled the coming together of disparate groups of stakeholders - including the authors of this article - to share their perspectives of what being a 'UKRI DRI stakeholder' meant for them in the context of 'net zero'. This collective sharing of a commonality, and the realization of the innate, potential power-for-good embodied within the community, catalyzed the research described in this article, namely devising and implementation of a series of stakeholder workshops. Their aims included giving voice to and empowering the UKRI DRI stakeholders to get involved in early planning and discussions on the actions necessary to achieve net zero; discovering potential best practices in engaging with the user community in an inclusive way around the net zero transition; and generating technical recommendations for the wider UKRI DRI scoping projects.

We designed a series of three workshops around predefined questions that were chosen from within the authors' accumulated stakeholder experience of UKRI DRI, within the more widely defined goal of achieving net zero by 2040. The workshop attendees were also UKRI DRI stakeholders and were aware of UKRI's net zero ambitions for its DRI. The findings from the workshops indicate that by the end of the series, the participants felt tangibly empowered, if not to take action within their job roles, then at least in their knowledge of their own potential impact on UKRI's journey to net zero.

Engaging with this user community lead to learnings around diversity and inclusivity: the participants' diversity in terms of their age, gender, nationality, job role and employer organisation reflected the diversity (i.e., lack of) within the DRI stakeholders that believed this workshop series to be relevant to them. Participants from research councils with a history of computationally demanding research (which should include at least BBSRC and MRC) dominated, yet not all research councils were represented. With hindsight this lack of inclusivity could have been addressed by seeking DRI champions from each of the nine UKRI organisations who would then be responsible for disseminating the net zero message through their own organisation.

The UKRI employees that attended the workshops had clearly self-identified as UKRI DRI stakeholders and they generated a rich source of technical recommendations for the UKRI DRI report. It is expected that these recommendations will contribute to UKRI's top-down policy, providing evidence that engaging with communities empowers them to design and plan a path to net zero. In the short-term future, receiving their own advice and recommendations framed as top-down policy should be further empowerment for the attendees.

Acknowledging the limitations of this study, we consider the presentation of predefined questions and how this shaped the workshop content, and the potential for the attendees to contribute. The workshop questions were stated in the advert, which could have excluded some applicants based on their own interpretations and understanding of the addressed topic. For future workshops we recommend a broader and more inclusive approach by holding a preliminary discussion across UKRI to define and refine the range of the DRI-related themes relevant to the whole community. The findings would determine the number, content, and format of the workshops necessary to represent the diversity of stakeholders across UKRI. This 'seed scattering' approach of holding a preliminary discussion should have the added advantage of attracting a larger audience to participate in the series of workshops, however they become defined. In turn, this would produce more robust statistical analyses.

A considerable challenge for many community-building activities is how to embed legacy to maintain and build on the momentum-for-action gained during the activity itself. In our case, limitations on funding and time meant we could produce a website and this article as follow-up to the workshops. Ideally, we would hold further community-gathering activities to encourage and support the attendees' planned actions towards net zero. For example, the vision piece describing "A day in the life of" a UKRI DRI user in the near future could be an effective method - given the powerful effect storytelling has on human society - to enable UKRI stakeholders to manifest their own actions towards net zero.

Our series of workshops was only a first step along the road to UKRI's ultimate goal of achieving net zero for its digital research infrastructure. The workshops have hinted that what the UKRI DRI stakeholders *could* achieve, given funding, time, and top-down 'permission', *could* be transformative, but there is a gap that needs to be bridged. The intention-to-action does not yet match the knowledge of the UKRI stakeholders about the effective action they could take. It would be interesting to learn if institutional support could bridge this gap and turn the previous '*coulds'* into '*woulds'*. Our final recommendation remains that system-level change will be necessary and must be driven from the community with support from the institutions in power.

**Conflict of Interest**

The authors declare that the research was conducted in the absence of any commercial or financial relationships that could be construed as a potential conflict of interest.

**Author Contributions**

Conceptualization: FA, DG, NM, JB, DM; Data curation: FA, DG, NM, JB, ALL, HK, NF, HH, DM; Formal Analysis: FA, DG, NM, JB, DM; Funding acquisition: FA, DG, NM, JB, ALL, DM; Investigation:  FA, DG, NM, JB, ALL, HK, NF, HH, DM; Methodology: FA, DG, NM, JB, DM; Project administration: DM; Supervision: DM; Visualization: NF; Writing – original draft: FA, DG, NM, JB; Writing – review & editing: FA, DG, NM, JB, DM

**Funding**

The authors received funding from NERC (Reference: NE/W007134/1).

**Acknowledgments**

The authors would like to thank the workshop participants, the Net Zero UKRI DRI Scoping team, and the reviewers of this paper.

# 7  Supplementary Material

The supplementary material comprise the devised action plans and the developed workshop toolkit. Anonymised workshop transcripts from participants who consented to data sharing are available in the supplementary material. The supplementary material can be accessed on request.

# 8  Appendix

Table 1 Participant information based on pre-workshop 1 survey from the EOI (reproduced from the original project report (Manika, Danae et al., 2022))

| Council Affiliations for Participants* | | Areas of Interest* | | Gender | |
|---|---|---|---|---|---|
| Councils | Number of participants | Area | Number of participants | Males | 14 |
| Arts and Humanities Research Council (AHRC) | 1 | HPC Infrastructure | 12 | Females | 10 |
| Biotechnology and Biological Sciences Research Council (BBSRC) | 2 | Data Centre | 17 | Age Category | |
| EPSRC | 5 | Software Development | 16 | 21-30 | 4 |
| Economic and Social Research Council (ESRC) | 3 | Hardware | 8 | 31-40 | 4 |
| Innovate UK | 1 | Training Skills | 11 | 41-50 | 10 |
| Medical Research Council (MRC) | 0 | Other | 3 (all the above, Research data and archiving) | 51-60 | 5 |
| NERC | 9 | Highest Education Level | | 61+ | 0 |
| Research England | 1 | High school diploma | 1 | Prefer not to disclose/Missing | 1 |
| STFC | 6 | Undergraduate degree | 4 | Country of Residence | |
| Central UKRI | 1 | Master's degree | 5 | UK | 20 |
| None of the above | 8 | PhD ongoing | 1 | China | 1 |
| Other | University funding, BEIS, Code carbon, Digital Humanities Climate Coalition, Software | PhD degree | 13 | France | 2 |
| | Development Organization | | | | |
| **Job Role** | | **Years in current role** | | Poland | 1 |
| Junior | 6 | < 5 years | 9 | **Incentive Choices** | |
| Senior | 16 | 5-10 years | 7 | Ecology donation (altruistic incentive connected to net zero/project) | 6 |
| Top Management | 1 | 10-15 years | 1 | UNICEF donation (altruistic incentive not connected to net zero/project) | 11 |
| Other | 1 (e.g., Junior pay but senior responsibility) | 15+ years | 7 | Lifestyle Voucher (Egoistic incentive) | 9 |

Notes: * participants could select more than one category; Total may or may not add up to 24 in total, due to some missing information from participants.

Table 2 Survey results before/after workshops on a 10-point likert scale. Please note that number of individual responses the surveys varied between workshops, and does not necessarily represent the total number of participants in each workshop.

| Construct | Pre-workshops survey/EOI M (SD) N=24 | After workshop 1 survey M (SD) N=29 | After workshop 2 survey M (SD) N=16 | After workshop 3 survey M (SD) N=10 |
|---|---|---|---|---|
| How would you evaluate workshop X overall? | n/a | 8.45 (1.53) | 8.50 (0.82) | 8.70 (0.95) |
| How comfortable do you feel about engaging with UKRI DRI stakeholders? | 8.25 (1.51) | 7.83 (2.16) | 8.25 (1.53) | 9.00 (1.55) |
| How important do you think it is for UKRI DRI becoming net zero by 2040? | 9.63 (0.92) | 9.59 (0.73) | 9.75 (0.58) | 9.80 (0.42) |
| How would you rate the importance of net zero in your job role? | 8.63 (1.35) | 8.03 (2.13) | 8.75 (1.06) | 9.20 (0.79) |
| How would you rate your current knowledge of developing your own work-related net zero action plan? | 5.92 (2.34) | 6.66 (2.16) | 7.31 (1.66) | 8.70 (1.16) |
| How much control do you think you have in terms of achieving net zero by 2040 in your job? | 6.29 (1.94) | 5.48 (2.05) | 6.88 (1.31) | 7.00 (1.76) |
| How would you rate your intentions to develop your own work-related net zero action plan? | 8.50 (1.35) | 8.28 (1.75) | 8.94 (1.18) | 9.50 (0.71) |